\documentclass{PoS}
\usepackage{amsmath}
\usepackage{amsfonts}
\usepackage{amssymb}
\usepackage{lmodern,dsfont}
\usepackage{cite}
\title{An analysis of the Lattice QCD spectra for $D^*_{s0}(2317)$ and $D^*_{s1}(2460)$}

\ShortTitle{An analysis of the Lattice QCD spectra for $D^*_{s0}(2317)$ and $D^*_{s1}(2460)$}

\author{\speaker{A. Mart\'inez Torres}$^a$, E. Oset$^b$, S. Prelovsek$^{c,d,e}$, A. Ramos$^f$\\
\llap{$^a$}Instituto de F\'isica, Universidade de S\~ao Paulo, Rua do Mat\~ao 1371, Butant\~a, CEP 05508-090, São Paulo, S\~ao Paulo,Brazil\\
\llap{$^b$}Departamento de F\'isica Te\'orica and IFIC,Centro Mixto Universidad de Valencia-CSIC Institutos de Investigaci\'on de Paterna, Aptdo. 22085, 46071 Valencia, Spain.\\
\llap{$^c$}Instit\"ut f\"ur Theoretische Physik, Universit\"at Regensburg, D-93040 Regensburg, Germany. \\
\llap{$^d$}Faculty of Mathematics and Physics, University of Ljubljana, 1000 Ljubljana, Slovenia. \\
\llap{$^e$}Jozef Stefan,Institute, 1000 Ljubljana, Slovenia.\\
\llap{$^f$}Departament d'Estructura i Constituents de la Mat\`eria and Institut de Ci\`encies del Cosmos, Universitat de Barcelona, Mart\'i i Franqu\`es 1, E-08028 Barcelona, Spain.
E-mail: \email{amartine@if.usp.br}, \email{oset@ific.uv.es}, \email{sasa.prelovsek@ijs.si}, \email{ramos@ecm.ub.edu}
}

\abstract{In this talk I present the results obtained using effective field theories in a finite volume from a reanalysis of lattice data on the $KD^{(*)}$ systems, where bound states of $KD$ and $KD^*$ are found and associated with the states $D^*_{s0}(2317)$ and $D^*_{s1}(2460)$, respectively. We confirm the presence of such states on the lattice data and determine the weight of the $KD$ channel in the wave function of $D^*_{s0}(2317)$ and that of $KD^*$ in the wave function of $D^*_{s1}(2460)$. Our results indicate a large meson-meson component in both cases.}

\FullConference{XVII International Conference on Hadron Spectroscopy and Structure - Hadron2017\\
		25-29 September, 2017\\
		University of Salamanca, Salamanca, Spain}

\begin{document}
\section{Introduction}
Lattice QCD studies of hadron systems aim at determining the spectra of mesons and baryons. To accomplish this objective, one of the difficulties to face is the fact that hadron resonances can not be related directly to the energy levels obtained in the spectrum of the QCD Hamiltonian when discretizing the spacetime. In spite of such a difficulty, lot of progress has been made in this matter of concern and information about hadron resonances are obtained within the L\"uscher method~\cite{lu,dudek,morningstar}. Particularly, in Ref.~\cite{lang} three energy levels were obtained in a lattice QCD simulation of the scalar $KD$ system when using $KD$ and $\bar s c$ interpolators: $2086~\pm 34$ MeV, $2218\pm 33$ MeV and $2419\pm 36$ MeV. Similarly, three energy levels were found in Ref.~\cite{lang} in the lattice QCD simulation of the axial $KD^*$ system when considering $K D^*$ and $\bar s c$ interpolators: $2232\pm 33$ MeV, $2349\pm 34$ MeV and $2528\pm 53$ MeV. Using the L\"usher method, the phase shift for scalar and axial systems in an infinite volume were calculated in Ref.~\cite{lang}, and through the effective range formula, the binding energies of the scalar $D^*_{s0}(2317)$ and of the axial $D^*_{s1}(2460)$ were determined to be around 40 MeV with respect to the $KD$ and $K D^*$ thresholds.

In this talk we present our results for the scalar $KD$ and axial $KD^* $ systems from a reanalysis of the energy levels obtained in Ref.~\cite{lang} using effective field theories in a finite volume~\cite{mar1,doring,mar2,mar3,alba}. The basic idea of the method is to solve the Bethe-Salpeter equation in a finite volume by using a parameter dependent kernel. The value of the parameters are determined from a fit to the lattice data, in this case, of Ref.~\cite{lang}. Using this kernel when studying the same system but at infinite volume, poles of the scattering matrix can be found in the complex energy plane. These poles are associated with the states $D^*_{s0}(2317)$ and $D^*_{s1}(2460)$. Information about the nature of these states can be also determined from the couplings of the poles found to the different channels used when solving the Bethe-Salpeter equation. In this way, effective field theories provide a valuable predictive and analyzing tool, alternative to the L\"usher method, which can constitute a breakthrough in the problem of analyzing resonances/bound states in Lattice QCD studies. In fact, a more recent lattice simulation of the $KD$ and $KD^*$ systems has been performed in Ref.~\cite{bali} in which the L\"usher method as well as effective field theories were used to analyze their data and similar conclusions to the ones presented here were found.

\section{Formalism}
In a box of volume  $V=L^3$, the Bethe-Salpeter equation reads as~\cite{mar1,doring,mar2,mar3,alba}
\begin{equation}
\mathcal{T}(E,L)=[1-\mathcal{V}(E)\mathcal{G}(E,L)]^{-1}\mathcal{V}(E),\label{bse}
\end{equation}
where $E$ is the center of mass energy of the system. In Eq.~(\ref{bse}), $\mathcal{V}(E)$ is a matrix whose elements are the lowest order amplitudes describing the transitions between meson-meson channels coupled to $KD^{(*)}$. For a transition $i\to j$  (with $i$ and $j$ representing the two meson initial and final states, respectively) we parametrize $\mathcal{V}_{ij}(E)$ as
\begin{align}
\mathcal{V}_{ij}(E,\alpha,\beta)=\alpha_{ij}+\beta_{ij}(s-s_\text{th}), \quad s_\text{th}=(M_{D^{(*)}}+M_K)^2,\label{ker}
\end{align}
with $s=E^2$ being the Mandelstam variable and $\alpha_{ij}$, $\beta_{ij}$ parameters which are determined by fitting the energy levels of Ref.~\cite{lang}. The parametrization in Eq.~(\ref{ker}) is based on the amplitudes found in the study of the meson-meson interaction within effective Lagrangians at infinite volume~\cite{daniel1,daniel2}, 

In Eq.~(\ref{bse}), for a two-meson channel $i$, $\mathcal{G}_i(E,L)$ is the corresponding two-meson loop function in the box, which is given by
\begin{align}
\mathcal{G}_i(E,L)&=G_i(E)+\lim_{q_\textrm{max}\to\infty}\left[\frac{1}{L^3}\sum_{q_r}^{q_\textrm{max}}I_i(\vec{q}_r)-\int\limits^{}_{q<q_\textrm{max}}\frac{d^3q}{(2\pi)^3}I_i(\vec{q}\,)\right],~~\vec{q}_r= \frac{2\pi}{L}\vec{n}_r, ~~\vec{n}_r \in \mathbb{Z}^3\nonumber\\
G_i(E)&=\int\frac{d^3q}{(2\pi)^3}I_i(\vec{q}\,),\quad\quad I_i(\vec{q}\,)=\frac{\omega_{1i}(\vec{q}\,)+\omega_{2i}(\vec{q}\,)}{2\omega_{1i}(\vec{q}\,)\omega_{2i}(\vec{q}\,)\left[s-(\omega_{1i}(\vec{q}\,)+
\omega_{2i}(\vec{q}\,))^2+i\epsilon\right]}.\label{loop}
\end{align}
In Eq.~(\ref{loop}), $G_i$ is the two meson loop function in the infinite volume for the channel $i$, which is regularized within a cut-off $q^\prime_\text{max}$, and $\omega_{1i,2i}(\vec{q})=\sqrt{\vec{q}^2+m^2_{1i,2i}}$ is the on-shell energy of the mesons 1 and 2, respectively, constituting the channel $i$. Different values of $q^\prime_\text{max}$ produce changes in $G_i$ which can be reabsorbed in the parameters $\alpha_{ij}$ and $\beta_{ij}$ of the kernel [Eq.~(\ref{ker})] when fitting the lattice data. Thus, any reasonable value of $q^\prime_\text{max}$, typically of the order of 1000 MeV, can be used to regularize $G$ and the results obtained are basically independent of $q^\prime_\text{max}$.

The eigen-energies of the system in the box of volume $V=L^3$ are then calculated from Eq.~(\ref{bse}) by solving
\begin{equation}
\textrm{det}[1-\mathcal{V}(E,\alpha,\beta)\mathcal{G}(E,L)]=0\label{det}.
\end{equation}
The resolution of Eq.~(\ref{det}) for different values of $L$ and for the $KD^{(*)}$ systems gives rise to energy levels comparable to those of Ref.~\cite{lang} and, at the same time, determines the parameters $\alpha_{ij}$ and $\beta_{ij}$ through the fitting of the data in Ref.~\cite{lang}. Once these parameters are known, $\mathcal{V}(E,\alpha,\beta)=\mathcal{V}(E)$ and we can use this potential to solve the Bethe-Salpeter equation at infinite volume, which is
\begin{align}
T(E)=[1-\mathcal{V}(E) G(E)]^{-1} \mathcal{V}(E).\label{bse2}
\end{align}
Poles of the scattering matrix $T$ in the complex energy plane are related to resonances/bound states. The coupling of these poles to the different meson-meson channels used when solving Eqs.~(\ref{bse}) and (\ref{bse2}) can be calculated from the residue of the $T$-matrix and the following sum rule is verified~\cite{daniel3,hyodo}
\begin{align}
-\sum_i g^2_i\frac{dG_i}{ds}\Bigg|_{\text{pole}}=1-Z\label{g_i}.
\end{align}
In Eq.~(\ref{g_i}), $g_i$ represents the coupling of the pole considered to a meson-meson channel $i$. Each term inside the summation symbol corresponds to the probability of finding the meson-meson channel $i$ in the wave function of the state related to the pole. The $Z$ represents the probability of finding any other component in the wave function of the state different to the meson-meson channels considered. In this way, we can deduce information about the nature of the resonance/bound state obtained.

\section{Results}
\begin{figure} 
\begin{center}
\includegraphics[width=0.45\textwidth,clip]{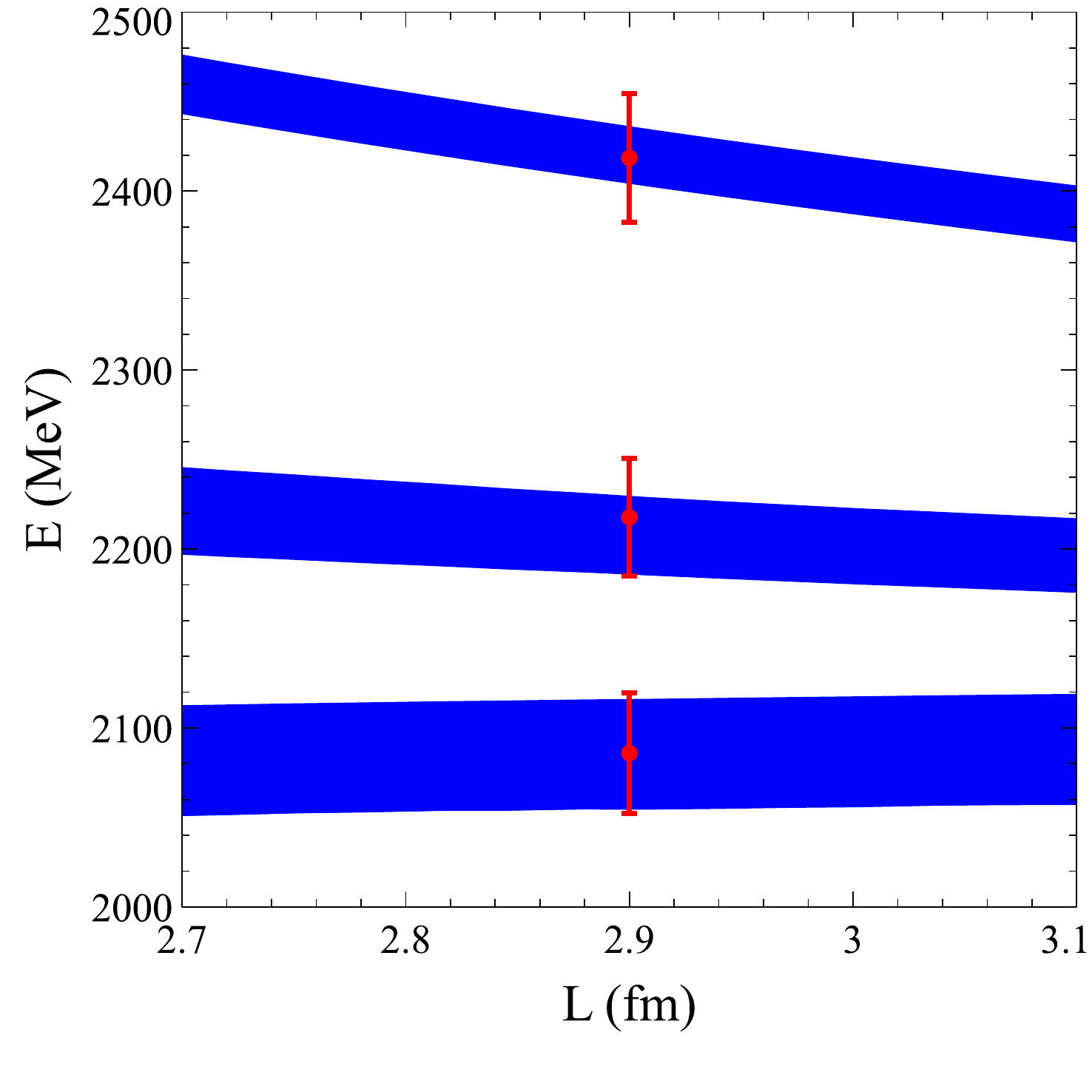}
\includegraphics[width=0.45\textwidth,clip]{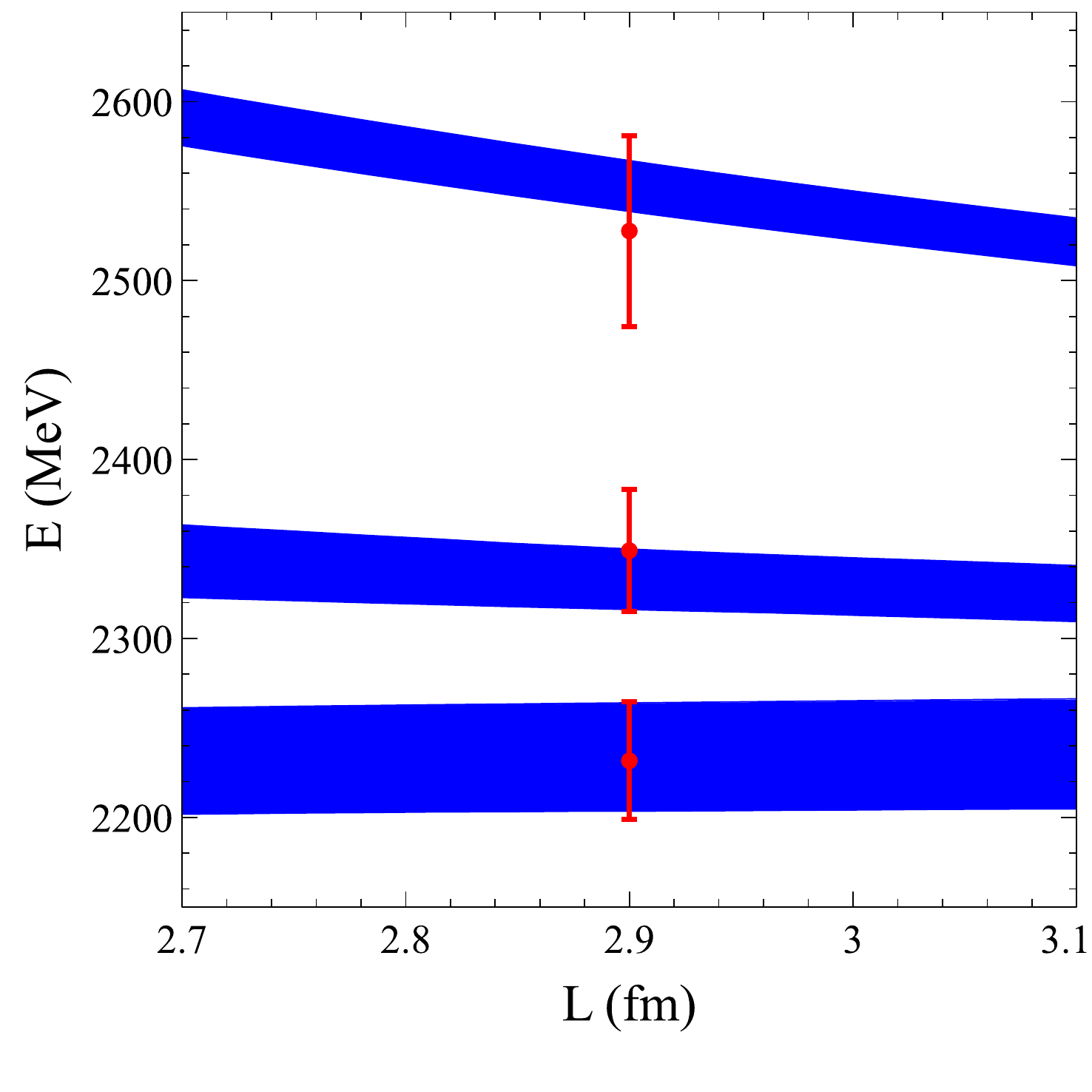}
\caption{Fits to the lattice data of Ref.~\cite{lang} for the $KD$ system (left panel) and the $KD^*$ system (right panel) which have been obtained when solving Eq.~(\ref{det}).}\label{res}
\end{center}
\end{figure}

In Fig.~\ref{res} we show the energy levels obtained from the fits to the data of Ref.~\cite{lang} when solving Eq.~(\ref{det}) considering $KD^{(*)}$ as the only coupled channel (in such a case, we just have the transition $KD^{(*)}\to K D^{(*)}$ and two parameters, $\alpha_{11}$ and $\beta_{11}$, need to be determined, where the subscript $1$ represents $KD^{(*)}$).

In the infinite volume case, the scattering matrix for the $KD\to KD$ transition reveals the presence of a pole whose binding energy with respect to the $KD$ threshold is found to be
$B(KD)=46\pm 21$ MeV and we can associate the pole with the state $D^*_{s0}(2317)$. This result can be compared with the one obtained in Ref.~\cite{lang} ( $B(KD)=36.6\pm 16.6\pm 0.5$ MeV) by means of the L\"usher method and the effective range formula. The probability of finding the $KD$ component in the wave function of $D^*_{s0}(2317)$ is found to be 
\begin{align}
76\pm 12~\%,\label{P1}
\end{align}
indicating that the state $D^*_{s0}(2317)$ has a large $K D$ component in its wave function. 

Similar is the case of the $KD^*$ channel, for which we find a pole with  
\begin{align}
B(KD^*)&=52\pm 22~\text{MeV},\nonumber\\
P(KD^*)&=53\pm 17~\%.\label{P2}
\end{align}
This pole can be related to the state $D^*_{s1}(2460)$.

We can also use the scattering matrix to determine the scattering length $a_0$ and the effective range $r_0$ and compare with the results found in Ref.~\cite{lang}. We find
\begin{align}
a_0(KD)=-1.2\pm 0.6~\text{fm},\quad r_0(KD)=0.04\pm 0.16~\text{fm},\nonumber\\
a_0(KD^{*})=-0.9\pm 0.3~\text{fm},\quad r_0(KD^*)=-0.3\pm 0.4~\text{fm},
\end{align}
which agree qualitatively well with those obtained in Ref.~\cite{lang}, yet we do not use the effective range formula.

In the effective field theories at infinite volume describing the $KD^{(*)}$ system and coupled channels, the existence of the states $D^*_{s0}(2317)$ and $D^*_{s1}(2460)$ is a consequence of the dynamics involved in the $KD^{(*)}$, $\eta D^{(*)}_s$ coupled channel system~\cite{daniel1,daniel2}, respectively. We might wonder about the relevance of the $\eta D_s$ channel in the wave function of $D^*_{s0}(2317)$ and that of the $\eta D^*_s$ channel in case of $D^*_{s1}(2460)$. With this idea in mind, we can solve Eq.~(\ref{det}) considering now $KD^{(*)}$ (associated with the label 1 below) and $\eta D^{(*)}_s$ (label 2) as coupled channels, in which case, 
\begin{align}
\mathcal{V}(E,\alpha,\beta)&=\left(\begin{array}{cc}\mathcal{V}_{11}(E,\alpha_{11},\beta_{11})&\mathcal{V}_{12}(E,\alpha_{12},\beta_{12})\\\mathcal{V}_{12}(E,\alpha_{12},\beta_{12})&\mathcal{V}_{22}(E,\alpha_{22},\beta_{22})\end{array}\right),\label{Vmat}\\
\mathcal{G}(E,L)&=\left(\begin{array}{cc}\mathcal{G}_{11}(E,L)&0\\0&G_{22}(E,L)\end{array}\right).
\end{align}
In Eq.~(\ref{Vmat}) we have used the fact that $\mathcal{V}_{21}=\mathcal{V}_{12}$. We have then 6 parameters to be determined by fitting the data of Ref.~\cite{lang}, but we have just three data points. In such a situation we can try to fit the data using energy independent kernels, i.e., 
\begin{align}
\mathcal{V}(\alpha)&=\left(\begin{array}{cc}\mathcal{V}_{11}(E,\alpha_{11},0)&\mathcal{V}_{12}(E,\alpha_{12},0)\\\mathcal{V}_{12}(E,\alpha_{12},0)&\mathcal{V}_{22}(E,\alpha_{22},0)\end{array}\right)=\left(\begin{array}{cc}\alpha_{11}&\alpha_{12}\\\alpha_{12}&\alpha_{22}\end{array}\right),\label{Valpha}
\end{align}
having in this way three parameters to be determined, $\alpha_{11}$, $\alpha_{12}$ and $\alpha_{22}$. By doing so, we would force to saturate the wave function of the states $D^*_{s0}(2317)$ and $D^*_{s1}(2460)$
with the $KD^{(*)}$ and $\eta D^{(*)}_s$ channels. This is so because, as shown in Ref.~\cite{hyodo}, the energy dependence of the kernel is related to the $Z$ function present in  Eq.~(\ref{g_i}). In this way, by comparing the probabilities found with just the $KD^{(*)}$ channel and an energy dependent kernel (Eqs.~(\ref{P1}) and (\ref{P2})) with those found with two coupled channels, $KD^{(*)}$ and $\eta D^{(*)}_s$, and constant kernel, we can obtain the weight of $\eta D_s$ in the wave function of $D^*_{s0}(2317)$ and that of $\eta D^*_s$ in the wave function of $D^*_{s1}(2460)$, which in Refs.~\cite{daniel1,daniel2} was found to be $\sim20\%$.

However, when trying to fit the data of Ref.~\cite{lang} by solving Eq.~(\ref{bse}) with the kernel in Eq.~(\ref{Valpha}), we do not find any suitable fit. This result could be interpreted as an evidence that the energy levels obtained in Ref.~\cite{lang} do not have information on the $\eta D_s$ or $\eta D^{(*)}_s$ channels:  although all states with a given quantum number are in principle expected in a dynamical lattice simulation, a poor basis of interpolating fields is insufficient to render them in practice. In this sense, the explicit consideration of $\eta D^{(*)}_s$ interpolators in a lattice simulation of the $KD^{(*)}$ systems should be considered in future to clarify the nature of $D^*_{s0}(2317)$ and $D^*_{s1}(2460)$.

\section{Conclusions}
In this talk we have presented the results found from a reanalysis of the lattice spectra obtained in Ref.~\cite{lang} for the $KD^{(*)}$ systems, where bound states were associated with $D^*_{s0}(2317)$ and $D^*_{s1} (2460)$. Our analysis confirms the existence of these bound states and the presence of a large $KD$ component in the wave function of $D^*_{s0}(2317)$ and $KD^*$ component in the wave function of $D^*_{s1} (2460)$. Our analysis suggests that future lattice simulations should explicitly include $\eta D^{(*)}_s$ interpolators, allowing in this way the determination of the probability of finding such components in the respective wave function of the states.

\section{Acknowledgements}
A.M.T gratefully acknowledges the financial support received from FAPESP (under  the
grant number 2012/50984-4) and the support from CNPq (under the grant number 310759/2016-1). This work is partly supported by the Spanish Ministerio de Economia y Competitividad and European FEDER funds under contract numbers FIS2011-28853-C02-01 and FIS2011-28853-C02-02,  by the Generalitat Valenciana in the program Prometeo II, 2014/068, and by Grant 2014SGR-401from the Generalitat de Catalunya. We acknowledge the support of the European Community-Research Infrastructure Integrating Activity Study of Strongly Interacting Matter (acronym
HadronPhysics3, Grant Agreement n. 283286) under the Seventh Framework Programme of EU.

\end{document}